# The Genetic Code via Gödel Encoding

**T. Négadi**

Physics Department, Faculty of Science, Oran University, 31100, Es-Sénia, Oran, Algeria; email: tnegadi@gmail.com; negadi_tidjani@univ-oran.dz

Abstract: The genetic code structure into distinct multiplet-classes as well as the numeric degeneracies of the latter are revealed by a two-step process. First, an empirical inventory of the degeneracies (of the shuffled multiplets) in two specific equal moieties of the experimental genetic code table is made and transcribed in the form of a sequence of integers. Second, a Gödel Encoding procedure is applied to the latter sequence delivering, as an output, a Gödel Number the digits of which, from the decimal representation, could remarkably describe the amino acids and the stops and allow us also to compute the exact degeneracies, class by class. The *standard* and the *vertebrate mitochondrial* genetic codes are considered and their multiplet structure is fully established.

## INTRODUCTION

Recently [1], we have published what was called by us an *observation* in a paper entitled *The genetic code multiplet structure, in one number*. The main result was the establishing of the correct multiplet structure of the *standard genetic code* into five amino acids degeneracy-classes and three stops, and the calculation of the degeneracy, in each class, using only the 23 decimal digits of the, at that time guessed, number 23!, as a starting point. It was also mentioned, briefly however, that the prime factorization of this latter number leads to **Ω**(23!)=41, where the completely additive arithmetic function **Ω,** also known as the big Omega function, counts the total *number* of prime factors including multiplicities, and 41 equals precisely the total degeneracy (61-20) for all five classes. In this letter, we precise this duality, make the number 23! *lawful*, embed it in a larger arithmetical setting by endowing it with the status of a *Gödel Number*, and show that the process of Gödel encoding allows us to establish a link, a "hinge", between the degenerated *codonic* part, encoded in the multiplicities of the prime factors, and the *amino acid* part, encoded in the decimal place-value representation. We also successfully extend our results to the *vertebrate mitochondrial genetic code*, the second most important version of the genetic code, after the standard one. Our method consists in partitioning the (experimental) genetic code table into two *particular* equal moieties and next apply the following two steps (i) make an empirical and *specific* inventory (see the last section) of the total degeneracy in each moeity, where the multiplets from different classes and the classes themselves are initially mixed, in the form of a decreasing sequence of integers and (ii) encode the latter sequence using Gödel encoding to obtain a Gödel Number. In terms of a set of digits, which we attach to the amino acids, and composing the decimal representation of said Gödel Number, the result of applying step (ii) gives, as an output, a new digital description of the amino acids classes and the stops. As a straightforward consequence, an easy computation of the detailed degeneracies for all the classes, from the digits themselves, is made possible. In the first section, we re-consider the case of the standard genetic code, in the light of our present formalization using the fundamental theorem of Arithmetic and Gödel encoding. In the second, we apply our method to the vertebrate mitochondrial code. The third and last section will be devoted to our concluding remarks and also to the presentation of some new striking relations, "etched" in the *atomic* content of the 20 amino acids and recording interesting numerical information. Before proceeding, let us mention a point on the degeneracy at the *first base-position* in the codon which is connected to the sextets, see [1]. There are two ways to view these latter: either as degeneracy-6 objects with 5 degenerate codons each or as separate degeneracy-2 and degeneracy-4 objects with respectively 1 and 3 degenerate codons. In the first case we have 20, as the number of amino acids (noted aas) whilst in the second we switch to what we may call Amino Acid Signals (AASs), which are identical to ordinary amino acids (aas) for 17 of them which are not degenerate, at the first-base position, and different only for the sextets $S^{II}$, $L^{II}$ , $R^{II}$, $S^{IV}$, $L^{IV}$ and $R^{IV}$ and we have now 23 as a new number representing the amino acids in the above generalized sense, i.e., 23 AASs. In the case of the vertebrate mitochondrial code, this number reduces to 22. Both views are described here.

## THE STANDARD GENETIC CODE

The genetic code is Life's dictionary that translates 64 triplet-codons into 20 amino acids and peptide-chain termination signals or stops. In the standard genetic code there are 61 meaningful codons for the amino acids and 3 stops. As several codons could code for one amino acid, one speaks about *degeneracy* and results a *multiplet structure* which is shown in Eq.(6). There are 5 quartets, 9 doublets, 3 sextets, 1 triplet and 2 singlets. A quartet is coded by four codons, a sextet by six, a doublet by two, a triplet by three and a singlet by one. Also, the multiplets are grouped into classes; in this case five. What we mean by degeneracy in the following is simply the number of codons in a multiplet *minus* one. For example a quartet has degeneracy 3, a doublet 1, a singlet 0, etc. Note that the total degeneracy for all five classes is $61-(5+9+3+1+2)=41=15+9+15+2+2\times0$. Consider now the two bisections of the standard genetic code table called by shCherbak, [2], 5'-pyrimidine and 5'-purine bisections, and shown in Table 1 below as the upper (lower) moieties, separated by a double line.

| $U^{(1)}$ | | | | $C^{(1)}$ | | | |
|---|---|---|---|---|---|---|---|
| F | F | S | S | L | L | P | P |
| L | L | S | S | L | L | P | P |
| Y | Y | C | C | H | H | R | R |
| s | s | s | W | Q | Q | R | R |
| I | I | T | T | V | V | A | A |
| I | M | T | T | V | V | A | A |
| N | N | S | S | D | D | G | G |
| K | K | R | R | E | E | G | G |
| $A^{(1)}$ | | | | $G^{(1)}$ | | | |

Table **1**: The standard genetic code table (Format from [8]).

We here note them $U^{(1)}+C^{(1)}$ ($A^{(1)}+G^{(1)}$) where the upper index means first-base in the codon. In the table the amino acids are written in one-letter code and s is for stops. From a biochemistry point of view, it is well known that the amino acids belonging to the *shikimate* and *glutamate* (*aspartate* and *pyruvate*) biosynthetic-pathway families tend to have codons belonging to $U^{(1)}+C^{(1)}$ ($A^{(1)}+G^{(1)}$) [9]. There are 11 AASs F, Y, C, W, H, Q, P, $S^{IV}$, $R^{IV}$, $L^{IV}$, $L^{II}$, 10 aas and 19 degenerate codons in $U^{(1)}+C^{(1)}$ and 12 AASs V, A, G, $S^{II}$, $R^{II}$, T, I, M, D, E, N, K, also 10 aas and 22 degenerate codons in $A^{(1)}+G^{(1)}$. Consider first $U^{(1)}+C^{(1)}$, *collectively* with its 19 degenerate codons, $A^{(1)}+G^{(1)}$, in some detail, and represent the total degeneracy as the following sequence:

$$S^{SC}: [19;9,4,3,2,1,1,1,1] \qquad (1)$$

For $A^{(1)}+G^{(1)}$, we have respectively three quartets with the same first base, V, A and G with 9 degenerate codons, two doublet-parts of two sextets $S^{II}$ and $R^{II}$ with 4 (2+2) degenerate codons, an isolated quartet T with 3 degenerate codons, the triplet I with 2 degenerate codons and finally four doublets D, E, N and K with a single degenerate codon each. Remind, that for methionine M the codon degeneracy is zero. This is our specific description of the two moieties. Some remarks on (and arguments for) this choice will be made in the last section. Now, let us apply step (ii) and encode the sequence (1) by using the well known Gödel encoding scheme [3][1]

$$enc(1)= 2^{19}\times\{3^9\times5^4\times7^3\times11^2\times13\times17\times19\times23\} \qquad (2)$$

The number in Eq.(2) is recognized as the prime factorization of the factorial of the number 23:

$$enc(1)=23!=25852016738884976640000 \qquad (3)$$

This last number was precisely our starting point in [1] as it was mentioned in the introduction but here it has been established as an output Gödel Number (the number enc(S) defined in the footnote is called a Gödel Number). Consider now the following set E constituted by the *23* decimal digits in *23*!

$$E : \{0,0,0,0,0,1,2,2,3,4,4,5,5,6,6,6,7,7,8,8,8,8,9\} \qquad (4)$$

A simple sorting logical procedure, described in [1], could be used to obtain the following digital pattern

$$\begin{array}{r l} 5 \text{ "quartets"} : & \{3,5,5,7,7\} \\ 9 \text{ "doublets"} : & \{4,4,6,6,6,8,8,8,8\} \\ 3 \text{ "sextets"} : & \{1,2,9\} \\ 1 \text{ "triplet"} : & \{2\} \\ 2 \text{ "singlets"} : & \{0,0\} \\ 3 \text{ "stops"} : & \{0,0,0\} \end{array} \qquad (5)$$

where the 23 digits could be associated, or attached, to the 20 amino acids and the three stops to match the structure of the standard genetic code into its five amino acid classes:

$$\begin{array}{r l} 5 \text{ quartets} : & \{ G, A, P, V, T\} \\ 9 \text{ doublets} : & \{ C, N, D, K, Q, E, H, F, Y \} \\ 3 \text{ sextets} : & \{S, L, R\} \\ 1 \text{ triplet} : & \{I\} \\ 2 \text{ singlets} : & \{M, W\} \\ 3 \text{ stops} : & \{\text{Amber, Ochre, Opal}\} \end{array} \qquad (6)$$

Note that 18 amino acids are coded by more than one codon and we have in Eq.(5) exactly 18 non-vanishing decimal digits. The 5 remaing zeros are attributed as follows: 2 for the two singlets M and W having codon-degeneracy zero and 3 for the three stops, as stop means "no amino acid" in protein synthesis. As a remark, there is some freedom in the digit-amino acid assignments *inside* each one of the three classes of even degeneracy but this is harmless and the situation is quite analogous to the one encountered when using group theory (see for example [4] where some amino acids assignments are defined up to permutations). Let us now proceed to the computation of some interesting numeric quantities from Eqs.(5). Denote by $\nu_i$ the number of digits and $\sigma_i$ the sum of their values, *without repetition*, in a given degeneracy-class i (1,2,3,4,6). We have

$$\begin{gathered} \nu_4 = 5,\ \sigma_4 = 15,\ \nu_4 + \sigma_4 = 20 \\ \nu_2 = 9,\ \sigma_2 = 18,\ \nu_2 + \sigma_2 = 27 \\ \nu_6 = 3,\ \sigma_6 = 12,\ \nu_6 + \sigma_6 = 15 \\ \nu_3 = 1,\ \sigma_3 = 2,\ \nu_3 + \sigma_3 = 3 \\ \nu_1 = 2,\ \sigma_1 = 0,\ \nu_1 + \sigma_1 = 2 \\ \nu_{st} = 3,\ \sigma_{st} = 0,\ \nu_{st} + \sigma_{st} = 3 \end{gathered} \tag{7}$$

[1] Let S= [$n_1$, $n_2$, $n_3$,…, $n_j$] be a sequence of integers. Then $enc(S)=2^{n_1}\times3^{n_2}\times5^{n_3}\times\ldots\times(\text{jth-prime})^{n_j}$.

In the above relations, we see at once that the five quartets, the two singlets and the unique triplet are perfectly described: $\nu$ gives the number of amino acids, $\sigma$ the number of degenerate codons and their sum the total number of codons. For the doublets and the sextets, some little work is needed as we shall see below but in the first instance, we could form the interesting total sum

$$(\nu_1+\sigma_1+\nu_3+\sigma_3+\nu_4+\sigma_4) + (\nu_2+\sigma_2+\nu_6+\sigma_6) = 67 \tag{8}$$

where classes 1, 3 and 4 are separated from classes 2 and 6. From Eqs.(7) and (8) we have 25+42=67. It has been shown in [1] that this is *exactly* the carbon-atom number distribution pattern in the 20 amino acids side-chains: 25 carbon atoms in classes 1, 3 and 4 and 42 in classes 2 and 6, with a total of 67. Let us now turn to the calculation of the degeneracies for the doublets and the sextets. For the former, there exist a fundamental symmetry, [5], implying the existence of a "one-to-one correspondence from one member of a doubly degenerate codon pair to the other member" and, in [1], we invoqued this fact to halve the $\sigma_2$-value. In practice, it suffices to introduce a coefficient, noted here $\alpha^{SC}$ and define a new $\sigma'_2=\alpha^{SC}\times\sigma_2$ in such a way that $\sigma'_2\equiv\nu_2$ because the number of codons is *equal* to the number of degenerate codons in doublets, i.e., 1 codon and 1 degenerate codon for any of them. We have readily $\alpha^{SC}=½$. The latter case requires counting the (number of) sextets two times, see at the end of the introduction, so that by taking $2\nu_6=3+3$ we have now two possibilities (i)

$$(3+3)+12=18 \tag{9}$$

or (ii)

$$3+(3+12)=18 \tag{10}$$

These relations describe correctly the two "views" the sextets could be seen. In the latter, we have 3 aas S, L and R and 15 degenerate codons whilst in the former we have 6 AASs $S^{II}$, $L^{II}$, $R^{II}$ and $S^{IV}$, $L^{IV}$, $R^{IV}$ and 12 degenerate codons. Moreover the three digits 1, 2 and 9 with sum $\sigma_6=12$ and attached to the three sextets (see Eq.(5)) have another very interesting virtue: they also record the degeneracy as 1 for $S^{II}$, 2 for $L^{II}$ and $R^{II}$ and 9 for $S^{IV}$, $L^{IV}$ and $R^{IV}$, for case (i). Below, we sum up the detailed results

$$\begin{gathered} \text{quartet}: \nu_4=5,\ \sigma_4=15,\ \nu_4+\sigma_4=20 \\ \text{doublets}: \nu_2=9,\ \sigma'_2=9,\ \nu_2+\sigma_2=18 \\ \text{sextets}: 2\nu_6=3+3,\ \sigma_6=12,\ 2\nu_6+\sigma_6=\nu_6+(\nu_6+\sigma_6)=18 \\ \text{triplet}: \nu_3=1,\ \sigma_3=2,\ \nu_3+\sigma_3=3 \\ \text{singlets}: \nu_1=2,\ \sigma_1=0,\ \nu_1+\sigma_1=2 \\ \text{stops}: \nu_{st}=3,\ \sigma_{st}=0,\ \nu_{st}+\sigma_{st}=3 \end{gathered} \tag{11}$$

Adding everything together, we obtain

$$23+38+3=64 \tag{12}$$

corresponding to (i) and

$$20+41+3=64 \tag{13}$$

corresponding to (ii). In Eq.(12) we have 23 AASs, 38 degenerate codons and 3 stops, where we have used the expression $2\nu_6+\sigma_6$ for the sextets. In Eq.(13) we have 20 aas, 41 degenerate codons and also 3 stops, where we have used this time the other expression $\nu_6+(\nu_6+\sigma_6)$. As a final remark, the choice made at the begining of this section relative to the two moieties, that is, $U^{(1)}+C^{(1)}$ collectively and $A^{(1)}+G^{(1)}$ in some detail, could be inverted to "see" this time $U^{(1)}+C^{(1)}$ in some detail. By using the permutational freedom in the prime factorization, we could re-write the number in Eq.(2) as

$$2^{19}\times7^3 \times \{3^9\times5^4\times11^2\times13\times17\times19\times23\} \tag{14}$$

Here the first part $2^{19}\times7^3$ codes for $A^{(1)}+G^{(1)}$, collectively, with 22 (=19+3) degenerate codons and the second part, between braces, codes for $U^{(1)}+C^{(1)}$ with 19 (=9+4+2+1+1+1+1) degenerate codons, in some detail. One has 9 degenerate codons for P, $R^{IV}$ and $S^{IV}$, 4 for $L^{IV}$ and F, 2 for $L^{II}$ and finally 4×1 for Y, C, H and Q (0 for W).

# THE VM GENETIC CODE

In the case of the vertebrate mitochondrial code (VMC, see Table 2 below), there are 4 stops, as 2 doublets, and 60 meaningful codons for the same 20 amino acids. There are here only three classes of even multiplets 6 quartets, 12 doublets and 2 sextets. Arginine is now a quartet, leaving its old doublet-part in favor of a doublet of stops, while isoleucine, tryptophane and methionine become three new doublets. The total degeneracy for all 3 classes is 60-(6+12+2)=40=18+12+10. Next consider also the same two bisections, as for the standard code in section 1.

$U^{(1)}$ $C^{(1)}$

| | | | | | | | |
|---|---|---|---|---|---|---|---|
| F | F | S | S | L | L | P | P |
| L | L | S | S | L | L | P | P |
| Y | Y | C | C | H | H | R | R |
| s | s | W | W | Q | Q | R | R |
| I | I | T | T | V | V | A | A |
| M | M | T | T | V | V | A | A |
| N | N | S | S | D | D | G | G |
| K | K | s | s | E | E | G | G |

$A^{(1)}$ $G^{(1)}$

Table **2**: The vertebrate mitochondrial genetic code table.

Here there are 11 AASs F, Y, C, W, H, Q, P, $S^{IV}$, $R^{IV}$, $L^{IV}$, $L^{II}$, 10 aas and 20 degenerate codons in $U^{(1)}+C^{(1)}$ and also 11 AASs V, A, G, $S^{II}$, T, I, M, D, E, N, K, 10 aas and 20 degenerate codons in $A^{(1)}+G^{(1)}$. In this symmetrical situation, it could be easily seen that *both* moieties are described by the *same* following sequence

$$S^{VMC}: [20;9,3,2,1,1,1,1,1,1] \qquad (15)$$

Taking $U^{(1)}+C^{(1)}$, collectively, with its 20 degenerate codons, we have for $A^{(1)}+G^{(1)}$, in some detail, three quartets with the same first base V, A and G and 9 degenerate codons, an isolated quartet T with 3 degenerate codons, the doublet-part of serine $S^{II}$ with 2 degenerate codons, and finally six doublets D, E, N, K, I and M with a single degenerate codon each. Taking now $A^{(1)}+G^{(1)}$, collectively with also 20 degenerate codons, we have for $U^{(1)}+C^{(1)}$, in some detail, three quartets with the same first base P, $L^{IV}$ and $R^{IV}$ with 9 degenerate codons, an isolated quartet $S^{IV}$ with 3 degenerate codons, the doublet-part of leucine, $L^{II}$, with 2 degenerate codons, and finally six doublets D, E, N, K, I and M with a single degenerate codon each. As in section 1 we apply now step (ii) by Gödel encoding the sequence (15)

$$\text{enc}(15)= 2^{20}\times\{3^9\times5^3\times7^2\times11\times13\times17\times19\times23\times29\} \qquad (16)$$
$$=3894589534689165312000$$

The Gödel Number in the third line of Eq.(16) has a priori no particular name, as it was the case for the standard genetic code, but it could nevertheless be successfully used to describe the vertebreate mitochondrial code, as we now show. A first sorting of the *22* digits in the above Gödel Number gives immediately eight even numbers 2, 4, 4, 6, 6, 8, 8, 8, three zeros and eleven odd numbers 1, 1, 3, 3, 3, 5, 5, 5, 5, 9, 9, 9. We have found the following final sorting

6 “quartets” : {4,4,6,8,8,8}
12 “doublets” : {0,1,1,3,3,3,5,5,5,5,9,9,9}
2 “sextets” : {2,6} (17)
2 (doublets) “stops” : {0,0}

adequate as it fits nicely the class structure of the vertebrate mitochondrial code:

6 quartets: {G, A, P, V, T, R}
12 doublets:{C, N, D, K, Q, E, H, F, Y, M, W, I} (18)
2 sextets :{S, L}
4 stops : {Amber, Ochre, AGR}

Here also the digit-amino acid assignments inside each classe are defined up to permutations. Note, in Eq.(18), that AGR corresponds to a doublet of codons, AGA and AGG, coding now stop codons and arginine reduces to a quartet (see above). Proceeding as for Eq.(7), we form the following quantities using Eq.(17)

$$\nu_4=6,\ \sigma_4=18,\ \nu_4+\sigma_4=24$$
$$\nu_2=12,\ \sigma_2=18,\ \nu_2+\sigma_2=30 \qquad (19)$$
$$\nu_6=2,\ \sigma_6=8,\ \nu_6+\sigma_6=10$$
$$\nu_{st}=2\times2,\ \sigma_{st}=0,\ \nu_{st}+\sigma_{st}=4$$

We see at once that the six quartets are perfectly described having 18 degenerate codons and 24 for their total number (6+18). For the twelve doublets and the two sextets, there is here, also, a little work to do and we proceed as for the standard genetic code in section 2. For the former, we introduce the coefficient $\alpha^{MC}$, define a new $\sigma'_2=\alpha^{MC}\times\sigma_2$ and demand that $\sigma'_2\equiv\nu_2$. This gives $\alpha^{MC}=2/3$ so that $\sigma'_2=12$. For the latter, counting them two times gives (i)

$$(2+2)+8=12 \qquad (20)$$

or (ii)

$$2+(2+8)=12 \qquad (21)$$

Here also the two digits 2 and 6 with sum $\sigma_6=8$ and attached to the two sextets (see Eq.(17)) record the degeneracy as 2 for $L^{II}$ and $S^{II}$ and 6 for $S^{IV}$ and $L^{IV}$, for case (i). Below, we sum up the results for the mitochondrial code

$$
\begin{array}{l}
6 \text{ “quartets”} : \ \nu_4=6,\ \sigma_4=18,\ \nu_4+\sigma_4=24 \\
12 \text{ “doublets”} : \nu_2=12,\ \sigma'_2=12,\ \nu_2+\sigma_2=24 \\
2 \text{ “sextets”}: 2\nu_6=2+2,\ \sigma_6=8, \\
\qquad 2\nu_6+\sigma_6=\nu_6+(\nu_6+\sigma_6)=12 \\
2 \text{ “doublet-stops”} : \nu_{st}=2\times 2,\ \sigma_{st}=0,\ \nu_{st}+\sigma_{st}=4
\end{array} \qquad (22)
$$

Note that, here, we have two doublets of stops so that there are 4 codons. Putting finally everything together we could write either

$$22+38+4=64 \qquad (23)$$

in case (i) or

$$20+40+4=64 \qquad (24)$$

in case (ii). These last two equations are the equivalent of Eqs.(12) and (13) in section 2. In Eq.(23) we have 22 AASs, 38 degenerate codons and 4 stops, using Eq.(20). In Eq.(24) we have 20 aas, 40 degenerate codons and also 4 stops, using Eq.(21). Note that, when considering case (i), the standard and the vertebrate mitochondrial codes share the same number of degenerate codons, 38.

## CONCLUDING REMARKS

The Arithmetic Model of the genetic code, presented in this letter, is based on the use of a mathematical encoding technique known as Gödel encoding, well known in Computer Science, and also on the related fundamental theorem of Arithmetic. The experimental data, in our case the shuffled degeneracies, are gathered in the form of a sequence of integers which is next Gödel encoded to deliver, as an output, a Gödel Number describing, through its digits the amino acids to which they are attached, in their respective classes, and permitting the calculation of the degeneracies class by class, from the digits themselves. Interestingly, note that by turning the way round one could start from the Gödel Number, carry out a prime factorization using the fundamental theorem of Arithmetic, and finds again the degeneracy sequence, encoded in the exponents of the prime factors. We successfully applied this mathematical procedure to the standard genetic code in the second section and to the vertebrate mitochondrial code in the third one. It is interesting to note, as a first remark, that the symbol “0” could play several roles, at the same time: it could be attached to the stops, where it could simply mean “nothing” as an “intuitive zero”, as well as to certain amino acids, as for methionine and tryptophan in the standard genetic code where it could mean “degeneracy=0”, as an ordinary numeric zero. In the case of the vertebrate mitochondrial code it is attached to one of the twelve doublets (see Eq.(17)), without being related, or even equal, to the degeneracy of the corresponding amino acid, as it was the case in the standard genetic code. The relevance of zero in the arithmetical approach to the genetic code has been recently emphasized by shCherbak [6]. Our second important remark concerns the specific choice, we made, for the partition of the genetic code table into two equal moieties, to be encoded (see the introduction). The fact that we choosed and started from “$U^{(1)}+C^{(1)}$, collectively, and $A^{(1)}+G^{(1)}$, in some detail” was, first and mostly, guided, *a posteriori*, by the particular and suggestive expression of the prime factorization of the number 23!, which was first discovered, as was mentioned in the introduction in connexion with reference [1] and, second, supported, by the remarkable physico-mathematical structure of $U^{(1)}+C^{(1)}$ (over $A^{(1)}+G^{(1)}$), as we now briefly explain. First, it was shown by shCherbak, some years ago, that there exist, as one from several ones, a nice nucleon-number balance in $U^{(1)}+C^{(1)}$ (not in $A^{(1)}+G^{(1)}$) between the side-chains and the blocks of the amino acids. Second, the distribution of the hydrogen atoms in the side-chains of the amino acids using the totality of the (standard) genetic code table, or all the “61 amino acids”, is also remarkable for the first set $U^{(1)}+C^{(1)}$ as it practically “talks” about itself, not so much for the second $A^{(1)}+G^{(1)}$. As a matter of fact there are 358 hydrogen atoms in “61 amino acids” of which 190 are in $U^{(1)}+C^{(1)}$ and 168 in $A^{(1)}+G^{1)}$ (see below). As there are 10 amino acids (aas) in each set, the number 190, written as 19×10, appears as if it was storing the product of the number of degenerate codons, 19, and the number of amino acids, 10, see section 2. Even the sum 10+19=29 gives, as required, the number of *meaningful* codons in $U^{(1)}+C^{(1)}$; this is not the case for 168 in $A^{(1)}+G^{(1)}$. Finally, we end this letter with the presentation of some few new remarkable physico-mathematical relations that could be infered from the nucleon content, more precisely from the hydrogen atom content of the amino acids side-chains, and having something to do with the mathematical structure of the standard genetic code. More will be given in a following publication. To proceed, we shall use another interesting completely additive arithmetic function, $a_0(n)$, which gives the *sum* of the prime factors in the prime factorization of the integer n, counting the multiplicities, to extract useful numeric information. Consider again the hydrogen atom content. There are 117 hydrogen atoms in the 20 amino acids side-chains (shCherbak’s borrowing made, see below), distributed over the five classes as follows: quartets 21, sextets 22, doublets 50, triplet 9 and singlets 15. The number 117 could be obtained simply as 204-87 where 204 is the total number of atoms and 87 the total number of carbon, nitrogen, oxygen and sulfur atoms (CNOS), in the 20 side-chains, see the table in [1]. Taking the same partition as in the remarkable pattern in Eq.(8), we have 117=45+72. From these numbers we have $a_0(45)=11$ and $a_0(72)=12$ with sum 11+12=23. This is exactly the number of AASs in $U^{(1)}+C^{(1)}$ and $A^{(1)}+G^{1)}$, respectively (see Table 1 and above). Note that if shCherbak's borrowing is not made, then the number of hydrogen atoms in the 20 amino acids side-chains becomes 1+117=118 and $a_0(118)=2+59=61$. This is also a highly interesting relation as it gives the total number of meaningful codons, 61, sum of 59 codons for which there exist a synonimous alternative and 2 codons for which there exist no synonimous alternative, the 2 singlets M and W. As we have briefly mentioned above there are exactly 358 hydrogen atoms, i.e., $21\times4+22\times6+50\times2+9\times3+15\times1$, in the 61 "amino acids" of which 190 are in $U^{(1)}+C^{(1)}$ and 168 in $A^{(1)}+G^{1)}$. The number 358 (a 61-gonal number!) is also very interesting: (i) it has as digits three Fibonacci numbers 3, 5 and 8 and (ii) it has $a_0(358)=2+179=181$. These relations seem hidding the mathematical pattern of the exceptional structure of proline and also the related famous "Egyptian Triangle", see [6]. From (ii), 2 is the first prime and 179 and 181 are two twin primes, respectively, 41st and 42nd so that in terms of the prime ranking index, we have 1+41=42 (or 42-1=41). This is shCherbak's "imaginary borrowing of one nucleon from proline's side-chain, 42-1, in favour of its block, 73+1", which warrants the existence of the numerous beautiful nucleon number balances. As for (i), 3, 5 and 8 appear to be the **3**rd, **4**th and **5**th from the Fibonacci series {1, 2, 3, 5, 8, 13, …}, to the extent that the "initial conditions" F(0)=0 and F(1)=1 are discarded. We see therefore here some manifestation of the "Egyptian Triangle", or the first Pythagorean triple {3, 4, 5}, see [6]. Even the coefficients $\alpha^{SC}$ and $\alpha^{MC}$ introduced for the doublets, in respectively sections 1 and 2, have values that appear as ratios of the first Fibonacci numbers 1/2, 2/3, 3/5, …. Finally, returning a last time to the number 117, we have $a_0(117)=19$ which could maybe refer to proline, "*it would indeed shine by its absence*", as there 19 true amino acids and one unique "imino acid", proline. Also, composing the two numbers, we have $a_0(117+a_0(117))=23$. In the detail, this last number could be explicited and put in the suggestive form 17+(2+2+2)=23 and this is exactly the AAS-pattern evoqued at the end of the introduction and playing an important role in this work. The relevance of the number 23 to the study of the genetic code has been emphasized many years ago by Gavaudan [7] but he saw it rather as 20+3 where 3 is the number of stops. In our approach, it is related to the degeneracy at the *first base-position* in the codon and to the three sextets serine, leucine and arginine.

The questions relative to the origin of the genetic code and *why this latter is just the way it is* are still the object of intense research. Recently in 2004, [10], Trevors and Abel wrote a paper entitled "*Chance and necessity do not explain the origin of life*" and, as the first sentence in their conclusion, they say that "New approaches to investigating the origin of the genetic code are required.". Some ten years ago Di Giulio already expressed "a certain lack of clarity in this field of research" [11]. The above authors evoque and discuss the many seminal attempts throughout the world, since the latter forty years, of leading scientists as Crick, Jukes, Woese, Wong, Yarus, Freeland, Di Giulio, Guimaraes, Schimmel, Ribas de Pouplana, shCherbak and many others to understand the genetic code origin (all their names and the complete references could be found in [10]). Trevors and Abel express however a lack of full satisfaction and leave the door open to other new approaches and ideas. As a matter of fact, among the people cited by the above authors, shCherbak, see [6], has clearly detected the presence of the decimal place-value numerical system as well as an *acting* zero inside the genetic code and also many beautiful unique summations. We think that our present Arithmetical Model, which relies also on the decimal numerical system, with also a zero playing distinct roles, agrees with shCherbak's work and goes along parallel lines of thought. We hope that this work could perhaps shed some light on other researches.

## ACKOWLEDGMENTS